# PENGEMBANGAN DOMAIN SPECIFIC LANGUAGE UNTUK PENGELOLAAN DATA WAREHOUSE

[1]Muhamad Taufan
[2]I Made Wiryana

[1,2] Program Magister Manajemen Sistem Informasi, Universitas Gunadarma
[1] m.taufan@gmail.com
[2] mwiryana.mobile@gmail.com

## *ABSTRAK*

*Upaya peningkatan kinerja pelayanan atas transaksi pada suatu bank dapat dilakukan dengan cara retensi data, mengurangi volume data di dalam database produksi dengan cara memotong data historis sesuai dengan aturan pada suatu bank ke data warehouse. Perancangan dan implementasi aplikasi Domain Specific Language (DSL) untuk penanganan data transfer pada data warehouse dibagi menjadi analisis leksikal, analisis sintaks, analisis semantik dan penghasil kode. Tiap bagian memiliki karakteristik yang berbeda untuk menghasilkan suatu perintah eksekusi. Telah dikembangkan suatu aplikasi dengan metode DSL yang bermanfaat mengurangi kesalahan penulisan perintah bagi user biasa (non-teknis) dalam melakukan pemindahan data. Dari pengujian menghasilkan keputusan metode transfer Oracle sesuai dengan ukuran skala data tertentu.*

***Kata Kunci****: DSL, Retensi data, Kompiler, Data warehouse*

## *ABSTRACT*

*Efforts to improve the performance of services on the transaction at a bank can be done by performing data retention, reduce the volume of data in the database production by cutting the historical data in accordance with the rules in a bank to a data warehouse. Design and implementation of applications Domain Specific Language (DSL) for handling the data transfer on the data warehouse is divided into lexical analysis, syntax analysis, semantic analysis and code generation. Each part has different characteristics to produce an executable command. Has been developed an application with the DSL method, which is beneficial to reduce the error of writing a command for a normal (non-technical) way to transfer data. From the test result in a decision Oracle transfer method according to the size scale of a particular data.*

***Keyword(s)****: DSL, Data Retention, Compiler, Data warehouse*

# PENDAHULUAN

Performa pelayanan atas suatu transaksi merupakan aspek *intangible* yang berdampak secara tidak langsung terhadap citra (*image*) bank. Salah satunya dengan membatasi volume data di dalam disk, data nasabah akan lebih cepat diakses yang berarti meningkatkan performa layanan. Pembatasan volume data pada database produksi, dan melakukan penyimpanan terhadap data lama ke dalam suatu database khusus yang secara fisik atau lojik terpisah dari database produksi dikenal dengan istilah *data retention* atau retensi (penyimpanan) data.

Di suatu bank daerah dengan jumlah nasabah yang cukup besar terdapat transaksi rekening koran dan buku besar (*general ledger*) dimulai pada tahun 2002 sampai sekarang (2012). Volume data dan transaksi yang semakin besar berkaitan terhadap performa proses batch dan proses backup database, proses batch: perhari memerlukan waktu sekitar 4 jam dan untuk akhir bulan memerlukan waktu sekitar 7 jam (bulan Juli tahun 2011), backup harian maupun bulanan memerlukan waktu sekitar 4 jam.

Retensi data dilakukan dengan tujuan utama meningkatkan performa layanan transaksi, Untuk pelaksanaan retensi data diperlukan aplikasi/*software* untuk proses transfer data dari database produksi ke data warehouse. Permasalahan yang terjadi ketika melakukan proses pemindahan histori data dari database produksi ke data warehouse yang dilakukan secara manual adalah dari sisi waktu dan untuk pemindahan data dengan ukuran yang besar, proses transfer masih tergolong lama dengan menggunakan metode Oracle *Data Manipulation Language*/DML (Query). Dibutuhkan metode oracle lain dalam transfer data dengan skala besar. Terdapat beberapa metode Oracle lain untuk proses pemindahan data, yaitu SQL *Loader* dan *Transport Tablespace*.

Masing-masing metode pada Oracle mempunyai inisialisasi dan karakteristik yang berbeda. Kecepatan transfer data tergantung dari skala data dengan kombinasi metode oracle yang dilakukan. Permasalahan sering timbul apabila pengguna tidak hafal opsi/perintah sehingga terjadi kesalahan dalam memberikan perintah. Kesalahan penulisan dalam perintah yang sederhana menghasilkan waktu yang lama.

Oleh sebab itu dibutuhkan sebuah aplikasi yang tidak dalam skala besar akan tetapi dapat fleksibel terhadap perubahan-perubahan yang terjadi. *Domain Specific Language* (DSL) atau "*little languages*" adalah bentuk terbatas dari bahasa komputer yang di rancang untuk penanganan masalah-masalah secara spesifik [1]. Penggunaan DSL pada aplikasi data transfer dari database produksi ke da-ta warehouse akan lebih efektif, karena ketiga metode oracle dibuat dalam satu DSL tanpa harus melakukan langkah-langkah atau inisialisasi dari masing-masing metode Oracle. Semua akan terparameter dan sangat fleksibel sesuai dengan kebutuhan client serta perkembangan teknologi pada Oracle. Sebagai penunjang dari proses pemindahan data, pada data warehouse untuk table yang berukuran besar akan di partisi berdasarkan bulan dan tahun, histori pada data warehouse disimpan semua histori transaksi dari awal tahun sampai sekarang.

Tujuan penelitian ini adalah membangun suatu solusi yang berguna untuk mempermudah dan mengurangi kesalahan pengguna dalam melakukan aktivitas yang berkaitan dengan pemindahan data serta fleksibel sesuai dengan kebutuhan pengguna berbasiskan pendekatan *language oriented programming*. Manfaat yang diharapkan dari penelitian ini adalah memberikan kontribusi dalam pendekatan lain untuk menyelesaikan masalah komputasi. Dan mempermudah pengguna dalam melakukan keputusan transfer data terutama untuk data yang bervariasi dan dapat dilakukan oleh pengguna biasa.

## TINJAUAN PUSTAKA

*Domain Specific Language* (DSL) adalah *programming language* atau *executable* dari bahasa tertentu yang menawarkan, melalui notasi dan abstraksi yang sesuai, kekuatan ekspresif yang terfokus pada, dan biasanya terbatas pada, beberapa masalah domain [2] *Domain Specific Language* (DSL) mengurangi upaya dalam pengembangan yang dibutuhkan ketika pemeliharaan untuk digunakan kembali dilakukan [3]. Sehingga DSL memiliki bentuk terbatas dari bahasa computer yang dirancang untuk penanganan masalah-masalah secara spesifik [1]. *Domain Specific Language* dibuat secara khusus untuk aplikasi domain bukan untuk tujuan umum sehingga menangkap dengan tepat hasil dari semantik

Ada beberapa pattern yang dapat digunakan dalam pengembangan DSL [4] yaitu keputusan (*decision*) dalam mendukung sebuah DSL baru ini biasanya tidak mudah, analisis masalah domain diidentifikasi dan domain knowledge secara eksplisit dan implisit dikumpulkan, disain dapat ditandai sepanjang dua dimensi ortogonal dan implementasi dipilih ketika sebuah (*excecutable*) DSL dirancang.

## METODE PENELITIAN

Penelitian ini dimulai dengan memahami masalah nyata yaitu yang akan dipakai menggunakan metode penelitian perkembangan untuk menyelidiki pola dan peru-rutan pertumbuhan dan atau pengubahan sebagai fungsi waktu. Data yang dipelajari berasal dari histori transaksi rekening koran dan *general ledger* dari suatu bank daerah. Pendekatan dalam membangun sistem adalah pendekatan *Rapid Prototyping*. Metodologi penelitian dilakukan dengan beberapa tahapan:

1. Analisis domain: Dalam tahapan ini di jelaskan permasalahan di ruang lingkup perbankan, pemilihan data, pemilihan metode transfer oracle dan sebab pemilihan metode DSL dalam pembuatan aplikasi. Dari tahapan analisis menghasilkan bahasa pemrosesan yaitu bahasa dan notasi yang dipilih berdasarkan kebutuhan.

2. Implementasi bahasa: Strategi dari implementasi untuk *patterns* yang cocok pada permasalahan di analisis domain.

## Analisis Domain

Di suatu bank daerah dengan pertumbuhan transaksi sekitar 9 juta sampai dengan 16 juta per tahun mengakibatkan proses batch dan backup mengalami waktu yang cukup lama sekitar 8 jam untuk harian dan 11 jam untuk bulanan, salah satu cara untuk meningkatkan performa pelayanan transaksi pada bank adalah dengan membatasi volume data pada database produksi dengan cara memotong data tersebut dan menyimpannya pada media lain yang terpisah. Tidak semua data dapat dipotong, data yang dibutuhkan sebagai data utama atau biasa dikenal dengan master file tidak dapat kita lakukan pemotongan.

Data yang dapat dipotong dan pindahkan ke media lain/database retensi adalah data transaksi rekening koran dan buku besar (*general ledger*) sejak tahun 2006 sampai tahun 2011. Permasalahan yang terjadi ketika melakukan proses pemindahan histori data dari database produksi ke data warehouse yang dilakukan secara manual oleh metode transfer Oracle yang menjadi kendala adalah dari sisi waktu dan untuk data yang berukuran besar diperlukan metode transfer Oracle yang berbeda.

Data Manipulation Language/DML (*Query*), SQL *Loader* dan *Transport Tablespace* merupakan metode transfer Oracle yang merupakan alat untuk pemindahan histori data. Masing-masing metode pada Oracle mempunyai inisialisasi dan karakteristik yang berbeda. Kecepatan transfer data tergantung dari skala data dengan kombinasi metode oracle yang dilakukan. Permasalahan sering timbul apabila pengguna tidak hafal opsi/perintah sehingga terjadi kesalahan dalam memberikan perintah.

Diperlukan aplikasi agar proses pemindahan dapat dilakukan secara mudah, fleksibel dan cepat sesuai dengan kebutuhan user. Dengan menggunakan metode DSL, semua inisialisasi metode oracle di atas tidak lagi didefinisikan satu persatu dan mengurangi kesalahan dari user dalam melakukan inisialisasi. Pengukuran pada penelitian ini dapat menghasilkan keputusan dalam pemindahan data menggunakan metode oracle dengan skala ukuran tertentu.

**Definisi bahasa**

Suatu bahasa dapat didefinisikan dengan berbagai cara dan notasi [5], pada penelitian ini dipilih beberapa bahasa dan notasi, yaitu:

- Sintaks merupakan perintah dasar yang harus atau boleh ada, yaitu PINDAH, SUMBER, TUJUAN, TABEL, TABEL2, TGL_AWAL, TGL_AKHIR, METODE, IGNORE

- Notasi merupakan tanda dari sintaks dan isian sintaks, yaitu [ Menunjukkan penanda awal dari isi sintaks, ] Menunjukkan penanda akhir dari isi sintaks, / Menunjukkan penanda database koneksi atau tanggal atau bulan atau tahun, @ Menunjukkan penanda password pada koneksi (sumber maupun tujuan)

- List merupakan kalimat/karakter statis, yaitu Y, T, QUERY, LOADER dan TRANSPORTTABLESPACE

Penggambaran bahasa menggunakan FSD (Finite State Diagram) dan tata bahasa menggunakan BNF (Backus-Naur Form sintaks)

**Finite state diagram**

Finite State Diagram (FSD) digunakan untuk menggambarkan secara umum dan khusus dari statement beserta *clause* dan identifiernya. Statement umum dari bahasa yang dikembangkan ini adalah sebagai berikut:

```
S1 <eol> S2 <eol> ..... Sn <eol>
```

Syntax
```
Statement := <PINDAH> <SUMBER_CLAUSE> <TUJUAN_CLAUSE> <TABEL_CLAUSE>
<TABEL2_CLAUSE> <TGLAWAL_CLAUSE> <TGLAKHIR_CLAUSE> <METODE_CLAUSE>
<IGNORE_CLAUSE> <EOL>
```

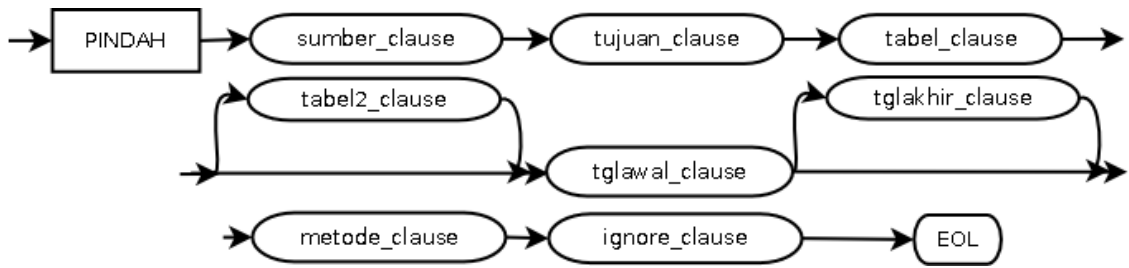

Gambar 1: FSD PINDAH

Statement ::=   Perintah umum dari transfer histori, statement dapat di input lebih dari satu. Masing-masing statement diakhiri dengan end of line. Finite State Diagram Statement terdapat pada Gambar 1

Contoh:

```
PINDAH SUMBER[..]  TUJUAN[..]   TABEL[..]    TGL_AWAL[..]  TGL_AKHIR[..]   METODE[..]
PINDAH SUMBER[..]  TUJUAN[..]   TABEL[..]    TGL_AWAL[..]  METODE[..]
PINDAH SUMBER[..]  TUJUAN[..]   TABEL[..]    TGL_AWAL[..]  METODE[..]  IGNORE[..]
PINDAH SUMBER[..]  TUJUAN[..]   TABEL[..]    TABEL2[..]   TGL_AWAL[..]  METODE[..]
```

**Spesifikasi Backus-Naur Form sintaks**

*Backus-Naur Form* (BNF) merupakan aturan yang menjelaskan sintaks dari bahasa yang dibuat. Berikut ini adalah notasi BNF yang digunakan:

Transfer *Statement*

```
<STATEMENT> ::= <PINDAH> <SUMBER> <TUJUAN> <TABEL> <TGL_AWAL> <METODE> <EOL> |
<PINDAH> <SUMBER> <TUJUAN> <TABEL> <TABEL2> <TGL_AWAL> <METODE> <EOL> |
<PINDAH> <SUMBER> <TUJUAN> <TABEL> <TGL_AWAL> <TGL_AKHIR> <METODE> <EOL> |
<PINDAH> <SUMBER> <TUJUAN> <TABEL> <TABEL2> <TGL_AWAL> <TGL_AKHIR> <METODE> <EOL> |
<PINDAH> <SUMBER> <TUJUAN> <TABEL> <TGL_AWAL> <METODE> <IGNORE> <EOL> |
<PINDAH> <SUMBER> <TUJUAN> <TABEL> <TABEL2> <TGL_AWAL> <METODE> <IGNORE> <EOL> |
<PINDAH> <SUMBER> <TUJUAN> <TABEL> <TGL_AWAL> <TGL_AKHIR> <METODE> <IGNORE> <EOL> |
<PINDAH> <SUMBER> <TUJUAN> <TABEL> <TABEL2> <TGL_AWAL> <TGL_AKHIR> <METODE> <IGNORE>
<EOL>
```

# Strategi Implementasi

Pendekatan implementasi yang digunakan untuk penanganan data transfer ini menggunakan *implementation patterns Compiler/Application Generator* karena DSL yang dikembangkan dapat digunakan kembali   dengan transfer data di  *environment* yang berbeda, contoh database yang berbeda,   atau *Operating System* yang berbeda.   DSL ini diimplementasikan dengan membuat compiler untuk DSL tersebut dengan menggunakan bahasa Python karena Kemampu-rawatan kompilator, dan kemampuan untuk dengan mudah menambahkan fitur bahasa baru.

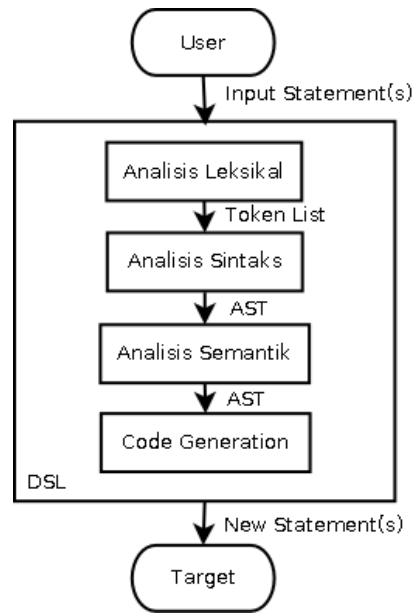

Gambar 2: Flow chart umum

Implementasi dari kompiler DSL yang dikembangkan memiliki sub-implementasi yang terdiri dari analisis leksikal yaitu aliran karakter yang mewakili *source* program dibaca dari kiri ke kanan dan dikelompokkan ke dalam token, analisis sintaks yaitu untuk menentukan apakah string token dapat dihasilkan oleh sebuah frase tata bahasa,  analisis semantik yaitu mengumpulkan informasi dan pengecekan terhadap data yang di masukkan tidak terdapat kesalahan semantik dan penghasil kode *(code generation*) yaitu mengumpulkan semua sintaks dari *Abstract Syntax Tree* menjadisebuah statement baru yang sudah ditentukan ulang dan dimapping sesuai kebutuhan untuk eksekusi transfer.

# HASIL DAN PEMBAHASAN

Penjabaran implementasi dari  kompiler DSL ini  akan dilakukan pada beberapa bagian yang penting yaitu, struktur bangun dari sistem kompiler DSL dan struktur data yang masing-masing terdiri dari analisis leksikal, analisis sintaks, analisis semantik dan penghasil kode (*code generation*) untuk digunakan di dalam kompiler DSL ini. Diagram blok struktur sistem secara umum terdapat pada Gambar 2

Flow chart tahapan analisis leksikal ini disajikan pada Gambar 3

Proses analisis leksikal, pendefinisian group token dan token dari input statement.

# Aturan / rules untuk analisis leksikal

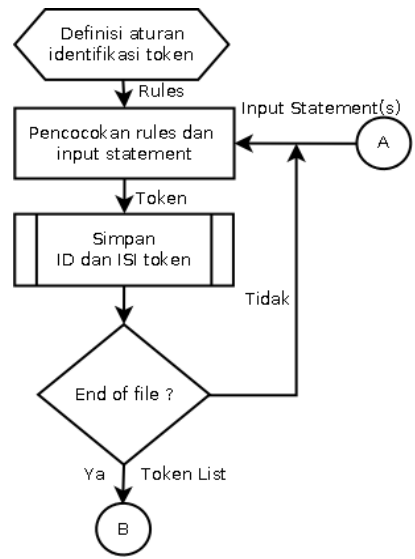

Gambar 3: Flow chart analisis leksikal

```
rules = [ ("KEYWORD", r"(PINDAH)"),

          ("IDENTIFIER", r"(SUMBER|TUJUAN|TABEL2|TABEL|TGL_AWAL|TGL_AKHIR|METODE|IGNORE)"),
          ("LITERAL", r"(LOADER|QUERY|TRANSPORTTABLESPACE)"),
          ("OPEN_SEPARATOR",r"[\[]"),
          ("CLOSE_SEPARATOR",r"[\]]"),
          ("OPERATOR",r"[\@]"),
          ("TGLBLNTHN",r"(\d|\d\d)\/(\d|\d\d)\/\d\d\d\d"),
          ("BLNTHN",r"(\d|\d\d)\/\d\d\d\d"),
          ("TAHUN",r"\d\d\d\d"),
          ("ALFANUMERIKSPESIAL",r"[\w$#]+"),
          ("STRING",r"(\"([^\"\]]+)\")"),
          ("SLASH",r"[\/\\]"),
          ("LINEBREAK",r"\n"),
          ("WHITESPACE",r"\s+"),
          ("END_STMNT", ("\n", stmnt_callback)),
          ]
# proses scanning statement dari aturan / rules yang ditetapkan
lex = Lexer(rules, case_sensitive=True)
for token in lex.scan(statement):
```

Flow chart analisis sintaks disajikan pada Gambar 4

Proses analisis sintaks, pendefinisian group token dan token dari input statement.

```
# Aturan untuk tata bahasa analisis sintaks

list_ident = ['PINDAH','SUMBER','TUJUAN','TABEL','TABEL2','TGL_AWAL','TGL_AKHIR',
'METODE','IGNORE']
map_ident = {} for i in range(0,9):
# pembentukan table rules analisis sintaks
if list_ident[i] in ('PINDAH'):
```

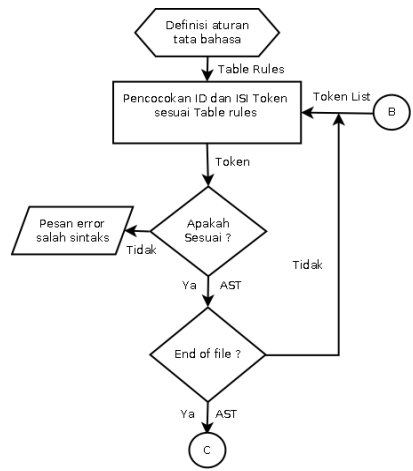

Gambar 4: Flow chart analisis sintaks

<span style="color:red"># setelah tanda [ terdapat komen, jika salah sintaks dan lanjutnya group</span>
<span style="color:red"># token/isi token yg benar</span>
map_ident[list_ident[i]] = ['"WHITESPACE must declared"|WHITESPACE|KEYWORD|
\s+']
map_ident[list_ident[i]].append('"IDENTIFIER must be declared"|IDENTIFIER')
..
..
<span style="color:red"># proses pengecekan analisis sintaks</span>
for i in range(0,len(id_token)):
    <span style="color:red"># jika token pertama bukan keyword, maka error</span>
    if inum==0 and (id_token[i]!='KEYWORD'):

        print('>    >>> Error line %d\n>>>>    >KEYWORD value must be declare, format
%s' %(lnum+1,id_token[i]))
        status=False
        break
    if id_token[i] in ('KEYWORD','IDENTIFIER'):
        for j in range(0,len(map_ident[isi_token[i]])):
          if i+j+1 < len(id_token):
          <span style="color:red"># jika group token tidak terdapat pada tabel rules maka sintaks error</span>
          if not id_token[i+j+1] in map_ident[isi_token[i]][j].split('|') or
not isi_token[i+j+1] in

            map_ident[isi_token[i]][j].split('|'):
            print('>>>> Error line %d\n>>>        >>%s %s
next value %s \nsintaks error %s wrong format %s '
%((lnum+1),id_token[i],isi_token[i],isi_token[i+j+1],
map_ident[isi_token[i]][j].split('|')[0],id_token[i+j+1]))
            status=False
            break
        ..

Flow chart analisis semantik disajikan pada Gambar 5
    Proses pengecekan terhadap analisis semantik

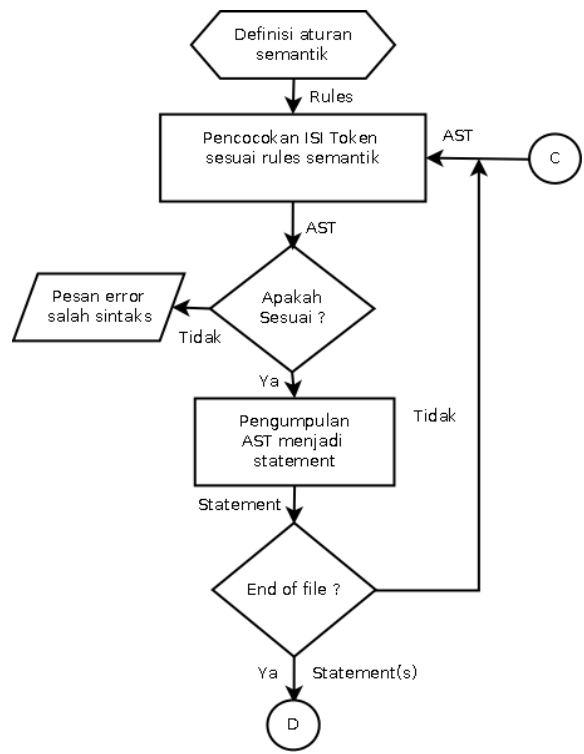

Gambar 5: Flow chart analisis semantik

<span style="color:red"># Cek AST (kelengkapan sintaks/kelebihan sintaks), pengecekan tanggal, transfor-</span>
<span style="color:red"># masi ke statement</span>
for i in range(0,len(id_token)):

    if id_token[i] in ('KEYWORD','IDENTIFIER'):

        hit_sintaks[isi_token[i]] = hit_sintaks[isi_token[i]] + 1

    <span style="color:red"># generate statement</span>
    try:

        statement[lnum]=statement[lnum]+isi_token[i].replace('\n',)
    except IndexError as info:
        statement.append(isi_token[i])
    if id_token[i] in ('LINEBREAK','END_STMNT'):
        for j in range(0,9):

            if hit_sintaks[list_ident[j]] > 1:

                <span style="color:red"># pengecekan semantic analysis kelengkapan sintaks, tidak</span>
                <span style="color:red"># boleh mendefinisikan lebih dari 1 sintaks dalam satu</span>
                <span style="color:red"># perintah (statement)</span>
                print('>>>   > Error Line %d IDENTIFIER %s tidak boleh lebih
                dari satu kali di definisikan \n' % (lnum+1,list_ident[j]))
                status=False
                status_line=status

            elif hit_sintaks[list_ident[j]] == 0:

                if (not list_ident[j] in ('TABEL2','TGL_AKHIR','IGNORE')):

```python
                        # pengecekan semantic analysis kelengkapan sintaks, semua
                        # sintaks mandatory ( TRANSFER, SUMBER, TUJUAN, TABEL,
                        # TGL_AWAL, METODE ) pada satu perintah (statement) harus
                        # didefinisikan
                        print('>>>   > Error Line %d IDENTIFIER %s belum di definisi-
                        kan \n' % (lnum+1,list_ident[j]))                  status=False
                        status_line=status
                else:

                        # assign nilai identifier yang boleh tidak didefinisikan
                        # oleh user, akan tetapi dibutuhkan oleh VM
                    if list_ident[j]=='TABEL2':
                    clause=statement[lnum].split(list_ident[j][:-1])[1].replace('[',)
                    clause=clause[:clause.find(']')]

                        TABEL2.append(clause)
                    elif list_ident[j]=='TGL_AKHIR':

                        clause=statement[lnum].split('TGL_AWAL')[1].replace('[',)
                        clause=clause[:clause.find(']')]
                        TGL_AKHIR.append(clause)
                    elif list_ident[j]=='IGNORE':
                        clause='T'
                        IGNORE.append(clause)
                    else:
                    # Untuk keperluan new statement nilai statement di transformasi
                    # sesuai kebutuhan VM
                    clause=statement[lnum].split(list_ident[j])[1].replace('[',)
                    clause=clause[:clause.find(']')]
                    if list_ident[j]=='SUMBER':

                        SUMBER.append(clause)
                ..
                ..
if status_line == True:
        # TGL_AKHIR tidak boleh lebih kecil dari TGL_AWAL
        if TGL_AKHIR[lnum] < TGL_AWAL[lnum]:

                print('>    >>> Line %d \nTGL_AWAL=%s cannot greater than
                TGL_AKHIR=%s\n' % (lnum+1,TGL_AWAL[lnum].strftime('%d/%m/%Y'),
                TGL_AKHIR[lnum].strftime('%d/%m/%Y')))
                status=False
                status_line=status
        else:
                TGL_AWAL[lnum]=TGL_AWAL[lnum].strftime('%d/%m/%Y')
                TGL_AKHIR[lnum]=TGL_AKHIR[lnum].strftime('%d/%m/%Y')

        # SUMBER dan TUJUAN boleh sama, selama nama TABEL tidak sama
        # dengan nama TABEL2
        if SUMBER[lnum]==TUJUAN[lnum] and TABEL[lnum]==TABEL2[lnum]:

                print('>    >>> Line %d \nTABEL=%s and TABEL2=%s cannot be
                same if SUMBER=TUJUAN \n' % (lnum+1,TABEL[lnum],TABEL2[lnum]))
                status=False
                status_line=status
```

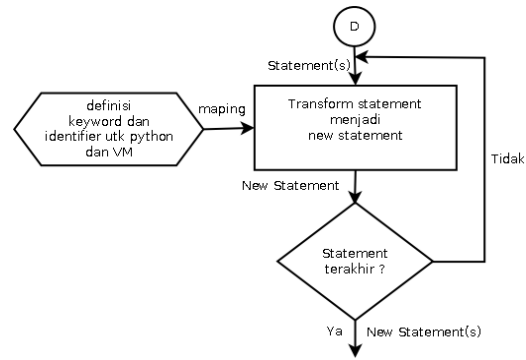

Gambar 6: Flow chart penghasil kode (*code generation*)

Flow chart tahapan penghasil kode (*code generation*) disajikan pada Gambar 6
Proses definisi maping dan proses transformasi dari statement ke statement baru.

```
# definisi KEYWORD dan IDENTIFIER utk python ke VM

map_codegen = [('PINDAH','TRANSFER'),('SUMBER','SOURCE'),
        ('TUJUAN','DESTINATION'),
        ('TABEL','TABLE'),('TABEL2','TABLE2'),('TGL_AWAL',
        'BEGIN_DATE'),('TGL_AKHIR','LAST_DATE'),
        ('METODE','METHOD')]

NEW_STATEMENT=STATEMENT
# proses transformasi dari STATEMENT ke NEW STATEMENT
for j in range(0,len(STATEMENT)):

        for i in range(0,len(map_codegen)):

                NEW_STATEMENT[j] = NEW_STATEMENT[j].replace
                (map_codegen[i][0],map_codegen[i][1])
```

## Pengujian

Pada pengujian dilakukan beberapa tahapan, yaitu pengujian fungsional, pengujian kinerja dan pengujian target. Pengujian fungsional dilakukan dengan cara memberikan masukan kepada blok dan melihat bagaimana blok melakukan penanganan dan memberikan keluaran. Dari pengujian fungsional tampak bahwa fungsi seluruh sistem saling terkait. Ketika salah satu blok dari sistem terdapat kesalahan, maka untuk blok selanjutnya tidak dapat dilanjutkan.

Pengujian kinerja (*performance*) merupakan tahapan untuk pengukuran kecepatan dari sisi kompiler dan dari sisi eksekusi pada bahasa target. Pengujian kinerja kompiler terdiri dari analisis leksikal, analisis sintaks, analisis semantik dan penghasil kode (*code generation*). Dari hasil kinerja (*performance*) test untuk proses kompiler, waktu proses paling lama dengan program berukuran besar

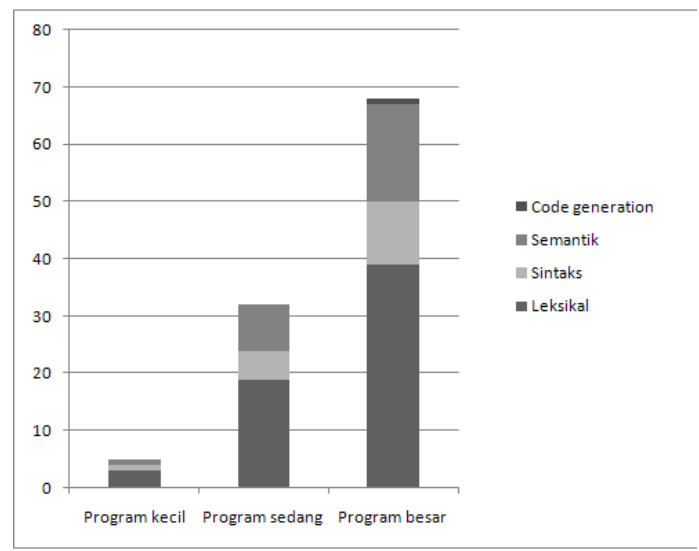

Gambar 7: Chart kompilasi

(lebih dari 100.000 LoC) adalah pada saat blok analisis leksikal (38 detik), kemudian blok analisis semantik (17 detik), lalu blok analisis sintaks (11 detik) dan yang terakhir adalah *code generation* (1 detik).

Dari hasil kinerja (*performance*) test untuk proses persiapan eksekusi pada target (*generate* sql/*execute* file), waktu proses paling lama dengan text ukuran besar (840 LoC) adalah pada metode transfer LOADER (30 menit), lalu TRANSPORT TABLESPACE (24 menit) dan yang terakhir QUERY (18 menit 50 detik). Waktu persiapan paling lama terdapat pada LOADER dan yang kedua adalah TRANSPORT TABLESPACE. Pada kedua metode tersebut terdapat beberapa inisialisasi yang harus dilakukan sebelum proses pemindahan data berlangsung. Sedangkan untuk QUERY lebih sederhana untuk proses inisialisasinya. Pada chart yang terdapat pada Gambar 7 menunjukkan bahwa analisis leksikal tetap stabil dalam mendominasi waktu proses kompilasi.

Pengujian target merupakan tahapan untuk pengukuran kecepatan hasil eksekusi dari kode yang dihasilkan. Terdiri dari proses kompilasi sampai eksekusi hasil kompilasi. Dari pengujian target untuk ukuran data histori yang kecil, proses pemindahan data yang paling cepat terdapat pada metode transfer Oracle QUERY disusul oleh SQL LOADER kemudian waktu yang paling lama adalah TRANSPORT TABLESPACE. Akan tetapi untuk ukuran data histori yang lebih besar, proses pemindahan data yang lebih cepat terdapat pada metode transfer Oracle TRANSPORT TABLESPACE disusul oleh QUERY lalu SQL LOADER. Pada chart yang terdapat pada Gambar 8 menunjukkan bahwa proses DSL stabil dengan kondisi normal akan tetapi jika terdapat kesalahan (dilakukan tanpa menggunakan DSL) maka akan

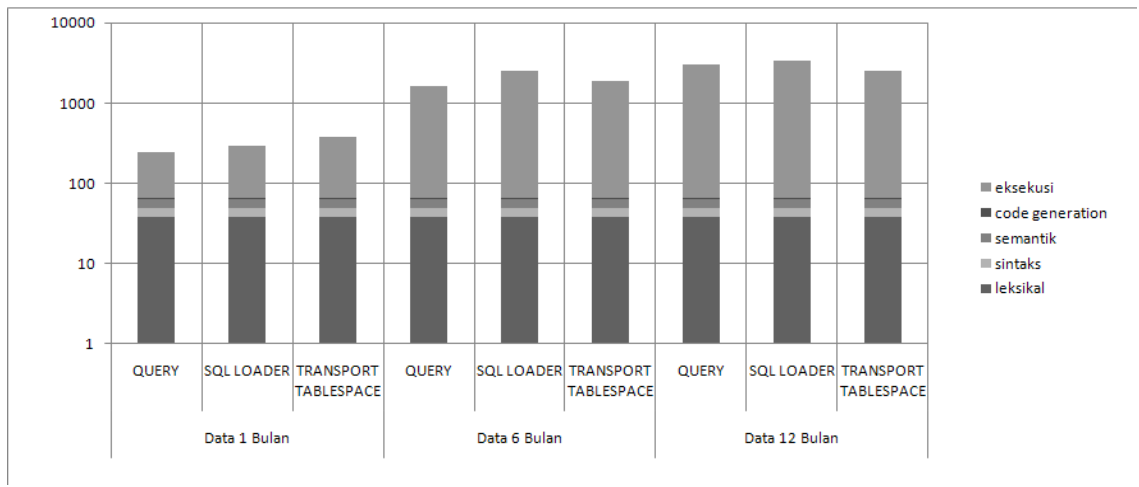

Gambar 8: Chart eksekusi target

sangat mempengaruhi dari waktu keseluruhan proses.

## Evaluasi

Dari penelitian ini telah dikembangkan DSL dengan teknik implementation patterns kompiler/*Application Generator* yang tediri dari analisis leksikal, analisis sintaks, analisis semantik dan penghasil kode (*code generation*). Dengan medefinisikan bahasa dan tata bahasa yang ditentukan untuk keperluan pemindahan data. Deskripsi dan definisi pemindahan data menjadi lebih mudah dan fleksibel sesuai dengan kriteria yang dibutuhkan user, bisa dilakukan dari satu database ke satu atau beberapa database lainnya, dengan nama tabel yang sama ataupun nama tabel yang berbeda, sehingga memudahkan user dalam melakukan interaksi dalam hal ini pemindahan data histori dari satu database ke satu atau banyak database.

Dari hasil pengujian kinerja ditunjukkan bahwa waktu kinerja kompiler yang paling lama terdapat pada blok analisis leksikal dimana blok tersebut merupakan tahapan awal dan fondasi bagi tahapan blok berikutnya. Jika terjadi kesalahan pada pendefinisian group token terutama pada urutan group token akan terjadi blok proses berikutnya yang tidak benar. Sedangkan untuk pengujian keseluruhan waktu pemindahan yang tercepat tergantung dari ukuran data yang akan dipindahkan, semakin kecil data yang dipindahkan metode Oracle QUERY merupakan pilihan yang terbaik. Akan tetapi untuk data yang berukuran besar perlu dipertimbangkan untuk menggunakan metode oracle TRANSPORT TABLESPACE

# KESIMPULAN DAN SARAN

Dari penelitian ini berhasil membangun solusi berbasiskan pendekatan *language oriented programming* dengan teknik implementation patterns kompiler/*Application Generator* yang berguna untuk pengguna biasa (non-teknis) dalam mempermudah dan mengurangi kesalahan pengguna melakukan aktivitas yang berkaitan dengan pemindahan data serta fleksibel sesuai dengan kebutuhan pengguna. Dari tahapan pengujian menghasilkan keputusan untuk mempermudah pengguna dalam melakukan keputusan transfer data terutama untuk data yang bervariasi.

Pada penulisan ini dari sisi perancangan DSL yang dikembangkan masih tergolong sederhana, disarankan untuk dapat menjadi lebih kompleks dan lebih banyak lagi untuk perintah dasar/*keyword*, sehingga user dapat melakukan interaksi dengan database menjadi lebih sederhana dan banyak hal dapat dilakukan tidak hanya pemindahan data, tapi misal membuat grafik, dan lain-lain. Patterns dari implementasi pada penulisan ini masih tergolong kaku untuk user dalam masukkan suatu instruksi/perintah, sehingga dapat dikembangkan untuk dapat dibuat dengan metode implementasi patterns *embeddeding* sehingga masukkan dari user dapat tidak dilakukan secara berulang, misal hanya membuat satu perintah/instruksi dengan kombinasi pengulangan (*loop*), maka dapat melakukan pemindahan data dari satu ke banyak database.

# Daftar Pustaka